\documentclass[a4paper,11pt]{article}
\pdfoutput=1 

\usepackage{jheppub} 

\usepackage[T1,T2A]{fontenc} 
\usepackage{graphicx}
\usepackage{dcolumn}
\usepackage{bm}
\usepackage{subfigure}
\usepackage{amssymb}
\usepackage{amsmath,amsfonts}
\usepackage[utf8]{inputenc}
\usepackage[english]{babel}
\usepackage{braket}

\newcommand{\beq}{\begin{equation}}
\newcommand{\eeq}{\end{equation}}
\newcommand{\beqar}{\begin{eqnarray}}
\newcommand{\eeqar}{\end{eqnarray}}

\newcommand{\bec}{\begin{center}}
\newcommand{\enc}{\end{center}}
\usepackage{color}

\title{\boldmath Unruh entropy of Schwarzschild black hole}

\author[a,b]{M.~Teslyk,}
\author[a]{L.~Bravina,}
\author[a,c,1]{E.~Zabrodin,\note{Corresponding author.}}
\author[b]{O.~Teslyk}

\affiliation[a]{
Department of Physics, University of Oslo,\\ PB 1048 Blindern,
Oslo N-0316, Norway}
\affiliation[b]{
Faculty of Physics, Taras Shevchenko National University of Kyiv,\\
Akademika Hlushkova Ave. 4, Kyiv UA-01033, Ukraine}
\affiliation[c]{
Skobeltsyn Institute of Nuclear Physics, Moscow State University,\\
Vorob'evy Gory, Moscow RU-119991, Russia}

\emailAdd{maksym.teslyk@fys.uio.no / machur@ukr.net}
\emailAdd{larissa.bravina@fys.uio.no}
\emailAdd{evgeny.zabrodin@fys.uio.no}
\emailAdd{teslyk.olena@knu.ua}

\abstract{
The entropy produced by the Unruh radiation is estimated and compared 
to the entropy of a Schwarzschild black hole. We simulate a spherical 
system of mass $ M $ by set of Unruh horizons and estimate the total 
entropy of the outgoing radiation. Dependence on mass and spin of the 
emitted particles is taken into account. The obtained results can be 
easily extended to any other intrinsic degrees of freedom of outgoing 
particles. The ratio of Unruh entropy to the Schwarzschild black hole 
entropy is derived in exact analytical form. For large black holes 
this ratio is highly sensitive to quantum numbers of emitted quanta, 
e.g., spin $ s $, for which it varies from $ 0\% $ for $ s = 0 $ to 
$ 19\% $ for $ s = 5/2 $. }

\begin{document}
\maketitle
\flushbottom

\section{Introduction}
\label{sec:introduction}


As known from the general relativity, any emission from a collapsing 
body having mass $ M $ gradually decreases while the matter falls 
inside down. Radiation from the surface vanishes completely after 
reaching some certain radius $ r = 2M $ in Planck units. After that 
any future light cone becomes directed inwards, and the collapsar 
transforms into a black hole. Being trapped below the 
event horizon, one needs a superluminal velocity to break the barrier. 
In relativity, a black hole is indeed black.
Surprisingly, such objects exhibit behavior typical for thermal systems. 
It turned out that a black hole has entropy proportional to its area 
\cite{bhi_Bekenstein}. Moreover, it obeys laws reminiscent of 
thermodynamics, namely \cite{bhi_4_laws}:
\begin{itemize}
	\item[0.] The surface gravity is the same all over the black 
hole's event horizon.
	\item[1.] The change of energy is expressed via the change of 
its area plus terms similar to work.
	\item[2.] The area of its event horizon cannot decrease.
	\item[3.] No finite process can eliminate the surface gravity 
completely.
\end{itemize}
The authors conclude that, as soon as the black hole cannot emit any 
radiation, its temperature should be zero. Therefore, one cannot expect 
the hole to be in equilibrium, because there is no two-way 
communication ordering to provide any kind of detailed balance. And 
this seemed to be an unsolvable mystery: how the black hole can have so 
much in common with a thermal system, if it cannot emit any radiation?

The solution came from an unexpected direction $-$ quantum physics 
\cite{bhi_Hawking}. Due to quantum fluctuations, particle-antiparticle 
pairs pop out from the vacuum even close to the horizon. One of them 
may be lucky enough to escape from a black hole, while its counterpart 
will fall inside. For any outer observer, the process looks as an 
outgoing thermal radiation. From the quantum point of view, a black 
hole is not black at all: it can emit radiation at the expense of its 
own energy.

Having solved one problem, the answer caused many others. As 
soon as the radiation from the event horizon is thermal, there is no 
chance for any information to escape. The emitted particles should be 
described not as a pure, but rather as a mixed quantum state 
\cite{PhysRevD.14.2460}. This is an essence of the information loss 
problem that questions the unitary of gravity.

The other issue is the amount of the black hole's entropy $H_{\rm BH}$. 
As shown by Hawking, it reads
\begin{equation}\label{bh:area law}
	H_{\rm BH} = A/4\ ,
\end{equation}
where $ A $ is the area of event horizon in Planck units. A simple 
estimate shows that even for the solar mass $ H_{\rm BH,\odot} \sim 
10^{77} $. So, what processes produce so much entropy? The question is 
not so simple: for a thermal system, the number of available 
microstates depends exponentially on its entropy. So, the puzzle 
becomes a real problem: what degrees of freedom are responsible for 
the area law, see Eq.\eqref{bh:area law}? To date, many approaches 
have been proposed to solve the problems.

The obvious suggestion was to count the relevant microstates. This was 
performed in different ways. Loop quantum gravity reproduces the area 
law by the quantization of black hole's phase space 
\cite{PhysRevLett.80.904,bhl_Khriplovich_1}. The result strongly 
depends on the exact value of the Barbero-Immirzi parameter 
\cite{bhl_e_spectr}, but its role still requires further clarification 
\cite{bhe_Immirzi}.

Alternatively, one can reduce the counting to the statistical problem 
from polymer physics \cite{bhl_su2,bhl_intertwiner}, thus assuming 
entropic origin of gravity \cite{bhef_Verlinde}. Similarly, the area 
law can be reproduced within the string theory by estimating the 
amount of string configurations 
\cite{bhe_str,bhe_micro,bhe_M-th,PhysRevD.75.084006,bhe_M-th_Kerr}.

Quantized black hole spectra offer an elegant solution to the problem
of information loss. Namely, particle evaporation should influence 
the dynamics of its internals under the horizon, thus resulting in 
less entropic radiation \cite{bhe_ev} and possible information 
outflow \cite{bhi_Hawking_2,1403.7314}.

It looks obvious that it is the event horizon that is responsible 
for the information loss and the entropy problems. This refers to 
such fruitful and interesting ways of solving a problem as the brick 
wall model \cite{bhe_Hooft,bhe_bw_unc,bhe_bw_cor}, firewalls 
\cite{Almheiri_2013,https://doi.org/10.1007/s10701-017-0122-3} or 
holography \cite{bhef_holo,bhef_en,bhef_en_ext,1108.2650}.

The event horizon of a black hole splits a whole space-time into 
accessible and inaccessible domains. Quantum mechanics states that 
one should take a partial trace over unobservant degrees of 
freedom, thus making the entanglement responsible for the entropy 
production at the horizon 
\cite{bhe_Srednicki,bhe_dof_rev,bhe_en_short,bhe_en_2,bhe_dof,bhe_atm}.

Despite that, none of the approaches has been widely accepted yet. 
Moreover, the necessity of an event horizon to reproduce thermodynamic
properties of a black hole can also be disputed 
\cite{mathur2023universality}.
The key problem is the poor knowledge of microscopic description for 
the space-time with horizons \cite{bht_rev,bht_rev_2}. Comprehensive 
discussions of the issues can be found in reviews 
\cite{1807.05864,1804.10610,Polchinski_2016,Unruh_2017,RevModPhys.88.015002}.

In present study we analyze the contribution of Unruh effect
\cite{u_Unruh} to the Bekenstein-Hawking entropy of a Schwarzschild 
black hole. A solely geometrical treatment for the entropy problem 
that origins from both peculiarities of space-time and Hilbert space, 
is an advantage of the approach.

The paper is organized as follows. Section~\ref{sec:probability} is 
devoted to some important issues of probability theory and information 
theory. Section~\ref{sec:unruh} briefly describes the Unruh radiation 
mechanism and the properties of its density matrix. Results of 
calculation of Unruh entropy that takes into account intrinsic degrees 
of freedom are presented in Section~\ref{sec:model}. 
Section~\ref{sec:analysis} deals with the analysis of asymptotics of
the obtained results. Details of the calculations are given 
in Appendix. Finally, conclusions are drawn in  
Section~\ref{sec:conclusion}. 

\section{Probability and entropy}
\label{sec:probability}

Let us consider some discrete non-normalized distribution $ \{X\} $, 
for which $ d\left(x\right) $ equals to the amount of events with $x$ 
being observed. The Shannon entropy $ H\left(X\right) $ for $ \{X\} $ 
may be written as
\begin{equation}\label{H(X)}
	H\left(X\right)
		=-\sum_{x}\frac{d\left(x\right)}{D_{X}}
			\ln\frac{d\left(x\right)}{D_{X}}
		= \ln D_{X} - \frac{1}{D_{X}}\sum_{x} d\left(x\right)\ln 
                       d\left(x\right)\ ,
\end{equation}
where $ D_{X} = \sum_{x} d\left(x\right) $. The entropy 
$ H\left(X\right) $ quantifies the information one needs to describe 
$\{X\}$, i.e., the amount of data we are lacking of about the system.

For a joint discrete distribution $ \{X,Y\} $ with the non-normalized 
distribution probability $ d\left(x,y\right) $ the situation looks 
similar. Its Shannon entropy $ H\left(X,Y\right) $ reads
\begin{equation}\label{H(X,Y)}
	H\left(X,Y\right)
		= \ln D_{X,Y} - \frac{1}{D_{X,Y}}
		\sum_{x,y} d\left(x,y\right)\ln d\left(x,y\right)\ ,
\end{equation}
where $ D_{X,Y} = \sum_{x,y} d\left(x,y\right) $.

At the same time, in the joint case one may introduce the conditional 
probability $ d\left(x|y\right) $ as
\begin{equation}\label{d(x|y)}
	d\left(x|y\right) = \frac{d\left(x,y\right)}{d\left(y\right)}\ , 
\qquad d\left(y\right) = \sum_x d\left(x,y\right)\ .
\end{equation}
It defines the fraction $ x $ from the subset of events with some 
certain value of $ y $. Using Eq.\eqref{H(X)}, the relevant Shannon 
entropy $ H\left(X|y\right) $ equals to
\begin{equation}\label{H(X|y)}
	H\left(X|y\right) = \ln D_{X|y} - \frac{1}{D_{X|y}}\sum_{x}
                           d\left(x|y\right)\ .
\end{equation}

Finally, substituting Eq.\eqref{d(x|y)} and Eq.\eqref{H(X|y)} into 
Eq.\eqref{H(X,Y)} one obtains
\begin{equation}\label{H conditional}
	H\left(X,Y\right)
		= H\left(Y\right)
		+ \Braket{H\left(X\vert y\right)}_{Y}
    	= H\left(X\right)
    	+ \Braket{H\left(Y\vert x\right)}_{X}\ ,
\end{equation}
where $ \Braket{\alpha}_{Y}=\sum_{y}\alpha d\left(y\right)/D_{Y} $.

It can be argued that the information entropy, defined by 
Eq.\eqref{H(X)}, differs significantly from that in thermodynamics. 
For any joint distribution describing correlated subsystems, the 
information entropy quantifies the amount of data encoded in these 
correlations. It should be proportional to the size of common 
boundary between the partitions. Therefore, the information entropy 
should be governed by some kind of an area law, see, e.g., 
\cite{RevModPhys.82.277}.

As known from thermodynamics, entropy is an extensive quantity which 
seems to contradict the conclusion above. To make both quantities 
compatible, one should take into account conditional distributions. 
In equilibrium any correlations vanish, so that the outcomes for 
$ x $ and $ y $ become independent. For Eq.\eqref{H conditional} one 
obtains then that $ H\left(Y|x\right) = H\left(Y\right) $, and the 
total entropy reads
\begin{equation}\label{H additive}
	H\left(X,Y\right) = H\left(X\right) + H\left(Y\right)\ ,
\end{equation}
i.e., additivity is restored.

This can be clearly seen for the Boltzmann case. Here momentum 
distributions $ \left\{X_i\right\}, i = \overline{1,N} $ of $ N $ 
particles are independent, and the total entropy 
$ H_{\rm B}\left(X_1,X_2,\dots,X_N\right) $ reads
\begin{equation}\label{H boltzmann}
	\left\{X_i\right\} \equiv \left\{X\right\}
	\quad\Rightarrow\quad
	H_{\rm B}\left(X_1,X_2,\dots,X_N\right)
		= N H\left(X\right)\ ,
\end{equation}
thus exhibiting bulk properties.

\section{Unruh radiation}\label{sec:unruh}
From here we will use Planck, or natural, units, i.e., $ G = c = 
\hbar = k_{\rm B} = 1 $.

Consider a spherically symmetric system of mass $ M $ and some 
quantum field surrounding it. The field is supposed to be in a pure 
vacuum state $ \Ket{0} $ in the free-falling reference frame and to 
have no influence on the background metric or the frame 
(quasiclassical approach). This condition implies that the field 
energy is negligibly small compared to $ M $.

Define an accelerated observer moving at acceleration 
$ \overrightarrow{a} $, with the norm $ \left|\overrightarrow{a}
\right| = a = \left(4M\right)^{-1} $. The corresponding non-inertial 
reference frame is small enough, so that one can neglect any tidal 
effects.

As was revealed by Unruh \cite{u_Unruh}, the definition for vacuum 
depends on the reference frame. In a curved space-time the emerging 
horizon splits a whole domain to the inside and outside partitions. 
Therefore, the state description will be different in each frame, 
depending on the preferred basis. For $ \Ket{0} $ it reads 
\cite{RevModPhys.80.787,0903.0250,0908.3149}
\begin{equation}\label{ket-0}
    \Ket{0} = \sqrt{\frac{1-e^{-E/T}}{1-e^{-NE/T}}}
    \sum_{n=0}^{N-1} e^{-nE/2T}\Ket{n}_{\rm in}\Ket{n}_{\rm out}\ ,
\end{equation}
where $ E $ is the energy of emitted quanta, and $ T $ is the Unruh 
temperature equal to 
$ T = \left(8\pi M\right)^{-1} = a\left(2\pi\right)^{-1} $. Parameter 
$ N $ determines the number of dimensions for the Hilbert space in a 
Fock basis. The ket-vectors with subscripts in the \emph{rhs} denote 
the corresponding Rindler modes with respect to the event horizon.

Regarding the physical meaning of $ N $, it stands for the largest 
number of emitted quanta at energy $ E $, which equals $ N - 1 $. For 
example, outgoing fermions restrict the value of $ N $ by 2. For 
bosons one usually sets $ N = \infty $ to have a complete Fock basis.
However, the correct value should obey the physical laws, including 
energy conservation. From this point of view, the assumption of 
infinite multiplicity is excessive, since no real physical system can 
emit an arbitrary number of particles \cite{particles5020014}. So, in 
what follows we assume that $ N $ is finite.

Expression \eqref{ket-0} is the Schmidt decomposition, see, e.g., 
\cite{pathak}, for which the density matrix of outgoing radiation 
reads
\begin{equation}\label{rho}
     \rho_{\rm out}
	= {\rm Tr}_{\rm in}\Ket{0}\Bra{0}
	=\frac{1-e^{-E/T}}{1-e^{-NE/T}}
	\sum_{n=0}^{N-1}e^{-nE/T}\Ket{n}_{\rm out}\Bra{n}_{\rm out}\ .
\end{equation}
Therefore, a pure vacuum state $ \Ket{0} $ transforms to the mixed 
one for the accelerated observer. From this it is easy to conclude 
that the only reason for the Unruh effect is geometry. Namely, the 
horizon arises from the finiteness of the speed of light and the 
absence of a preferred complete basis in the Hilbert space. The 
system as a whole is in a pure state and is governed by a unitary 
evolution. However, the imposed restrictions (horizon) make it 
impossible to monitor the global space, thus resulting to a mixed 
state for the radiation.

Each eigenvalue of $ \rho_{\rm out} $ quantifies the probability to 
detect $ n $ outgoing particles at energy $ E $ and temperature 
$ T $. From the information theory point of view, one deals with the 
conditional multiplicity distribution $ \left\{n|N,E/T\right\} $ at 
given $ N $ and $ E/T $. Its von Neumann entropy reads
\begin{eqnarray}\label{H_U(n|N,E/T)}
   H_{\rm U}\left(\rho_{\rm out}\right)
	= H\left(n|N, E/T\right)
	= \sigma\left(E/T\right) - \sigma\left(NE/T\right)\ ,
\end{eqnarray}
where
\begin{equation}\label{sigma}
   \sigma\left(qE/T\right) = \frac{qE/T}{e^{qE/T}-1} - 
                             \ln\left(1-e^{-qE/T}\right)\ .
\end{equation}

Quantity $ H\left(n|N, E/T\right) $ is an even function of $ E/T $. 
Its asymptotic behavior with respect to $ E/T $ is given by
\begin{equation}\label{H_U(n|N,E/T) asymptotics}
\begin{split}
	&\lim_{E/T\to 0}H\left(n|N, E/T\right)
		= \ln N
		= \max\left(H\right)\ ,  \\
	&\lim_{E/T\to\infty}H\left(n|N, E/T\right) = 0\ .
\end{split}
\end{equation}
This behavior can be explained as follows. At high temperatures, when 
the ratio $ E/T $ is small, the eigenvalues of $ \rho_{\rm out} $ 
approach a constant value. Physically it means that one can neglect 
any correlations induced by the energy conservation. It leads to 
homogeneous energy distribution of $ \left\{E\right\} $ at which the
entropy reaches its maximum. The other asymptotics applies to 
particle emission at energies that significantly exceed the source 
temperature. Being highly unlikely, such processes are exponentially 
suppressed. Therefore, the asymptotic behavior of
Eq.\eqref{H_U(n|N,E/T) asymptotics} is completely determined by the 
energy conservation.

\section{Model: basic features and results}
\label{sec:model}

The entropy $ H_{\rm U}\left(\rho_{\rm out}\right) $ from 
Eq.\eqref{H_U(n|N,E/T)} does not take into account any other degrees 
of freedom except the multiplicity $ n $ and energy $ E $ as a 
parameter. However, the emitted quanta may carry intrinsic degrees of 
freedom which should influence the phase space of radiation and, 
consequently, its entropy.

In addition, the contribution of relevant Hilbert subspace might 
induce additional conservation laws and affect the reference frame. 
For example, if some emitted particle carries out a non-zeroth spin, 
the angular momentum conservation dictates the source to change its 
background metric. Then, the distribution $ \left\{T\right\} $ over 
source temperatures should be analyzed also, thus significantly 
complicating any calculations. Besides, such extension will 
contradict the quasiclassical assumption, thus making the whole 
formalism questionable.

To overcome the problem, we suggest that the intrinsic degrees of 
freedom have no influence on the background metric, in full accord 
with the quasiclassical approach. This is valid for a large enough 
black hole, when imposing any new quantum number $ q $ causes 
negligible correlations. Thus, one may consider the emission 
probability to be independent of $ q $ -- similar to the analysis 
below, see Eq.\eqref{H_U(n|N,E/T) asymptotics}. So, due to 
Eq.\eqref{H additive}, any intrinsic degrees of freedom increase 
the Unruh entropy as

\begin{equation}\label{H(q,n|N,E/T)}
    H\left(Q,\rho_{\rm out}\right)
	= \ln D_{Q} + H_{\rm U}\left(\rho_{\rm out}\right)\ ,
\end{equation}
where $ D_{Q} $ is the number of dimensions of Hilbert space 
describing the relevant degree of freedom.

The Schmidt decomposition Eq.\eqref{ket-0} is defined for the 
$ D = 1 + 1 $ space-time. Any additional spatial dimensions can be 
omitted with no consequences for the density matrix $\rho_{\rm out}$. 
But any Schwarzschild black hole is embedded in a $ D = 3 + 1 $ 
space time. What is the contribution of the lower dimension effect, 
if any? To answer this, one should take angular degrees of freedom 
into account.

The Unruh effect has only one certain direction, which is determined 
by the unit vector $ \overrightarrow{a}/a $. It can be argued that 
the Unruh temperature $ T $ does not contain such information, since 
it is completely determined by the acceleration $ a $ 
\cite{PhysRevD.7.2850,10.1088/0305-4470/8/4/022,u_Unruh}:
\begin{equation}\label{T unruh}
	T = \frac{a}{2\pi}\ ,
\end{equation}
with no vector data inside.

The situation is similar to the black-body radiation. Despite the
fact that its emission spectrum carries no information about 
orientation, there is some other specific direction, which is 
determined by the momentum of emitted particles. For a $ D = 3 + 1 $ 
source, its radiation consists of similar sources, each generating 
$ H\left(q,\rho_{\rm out}\right)$. 
For the black body, one estimates its total entropy via the 
corresponding integral in the phase space. For a spherical shape, 
this makes entropy an extensive quantity, see the discussion at the 
end of Section~\ref{sec:probability}.

So, outgoing Unruh particles are emitted at some certain direction, 
which is encoded with $ \overrightarrow{a}/a $. Assuming that the 
total Unruh entropy $ H_{M} $ is extensive with respect to the 
angular degrees of freedom, we get
\begin{equation}\label{H_BH(s,n|N,E/T)}
    H_{M}\left(Q,n|N,E/T\right)
	= D_{\overrightarrow{a}/a}H\left(Q,\rho_{\rm out}\right)\ ,
\end{equation}
where $ D_{\overrightarrow{a}/a} $ denotes the number of 
distinguishable directions. Contrary to the phase space of 
black-body radiation, here we deal with density matrix. This means 
that $ D_{\overrightarrow{a}/a} $ is governed by eigenvalues $ l $ 
of angular momentum operator and its projection 
$ l_z, -l \leq l_z \leq l $.

As mentioned above, we are considering a spherically symmetric 
system. Any non-inertial observer will measure the outgoing radiation 
from the spherically-shaped domain having radius $ r = 2M $. For 
this, the angular momentum of the emitted particle is bounded as
\begin{equation}\label{l bounds}
	0 \leq \sqrt{l\left(l+1\right)} 
	  \leq \sqrt{L\left(L+1\right)}
	  = rp
	  = 2M\sqrt{E^2-m^2}\ ,
\end{equation}
where $ m $ is the particle mass and $ E $ is its energy. One might 
argue that $ L $ should be integer, which may not be the case for 
the \emph{rhs}  of Eq.\eqref{l bounds}. To overcome the problem, we 
assume that
\begin{equation*}
	2M\sqrt{\frac{E^2-m^2}{L\left(L+1\right)}}
		= 1 + \varepsilon\ , \quad
	\varepsilon \ll 1\ .
\end{equation*}

Therefore, the number $ D_{\overrightarrow{a}/a} $ is just the sum 
over all available $ l $ and $ l_z $:
\begin{equation}\label{D_a}
   D_{\overrightarrow{a}/a}
	= \sum_{l=0}^{l=L}\sum_{-l}^{l} = \left(L+1\right)^2
	= \frac{1}{4}\left(\sqrt{\frac{E^2-m^2}{4\pi^2T^2} + 1} 
            + 1\right)^2\ .
\end{equation}
Here we used Eq.\eqref{l bounds}.

Next, one should take the energy distribution $ \left\{E\right\} $ 
into account. In general case, the probability to emit a particle 
should be governed by the energy conservation. This means that the 
mass of black hole shrinks right after the particle leaves the event 
horizon. Any elementary emission affects the background metric due to 
the change of black hole's internal characteristics. Having no 
quantum gravity to extract the relevant information, we assume the 
entropy to be additive with respect to energy, similar to
Eq.\eqref{H boltzmann}.

We also suggest to limit the energy $ E $ of emitted quanta from 
above on the Planck scale
\begin{equation}\label{E limits}
	m \leq E \leq 1 \ll M\ .
\end{equation}

Therefore, the total Unruh entropy for the Schwarzschild black hole 
can be estimated as
\begin{equation}\label{H_BH(s,n,N,E|T)}
	H_{M}\left(Q,n,E|N,T\right)
		= \int_{m}^{1} D_{\overrightarrow{a}/a} \ln D_{Q} {\rm d}E
		+ \int_{m}^{1}
			D_{\overrightarrow{a}/a}
			H_{\rm U}\left(\rho_{\rm out}\right)
			{\rm d}E\ ,
\end{equation}
where we have used Eq.\eqref{H(q,n|N,E/T)}, and 
$ D_{\overrightarrow{a}/a} $ is determined by Eq.\eqref{D_a}.

Quantity $ H_{M}\left(Q,n,E|N,T\right) $ estimates the entropy produced 
by the joint distribution $ \left\{Q,n,E|N,T\right\} $ of Unruh 
radiation from the Schwarzschild black hole horizon having temperature 
$ T $. Recall that $ N $ determines the possible number of particles 
and is governed by both spin statistics and energy conservation.

Finally, the ratio of the Unruh entropy to the black hole entropy 
$ H_{\rm BH} = \left(16\pi T^2\right)^{-1} $ reads
\begin{equation}\label{ratio}
	\frac{H_{M}\left(Q,n,E|T\right)}{H_{\rm BH}}
		= \varUpsilon_{Q} + \varUpsilon_{\rm U}\ ,
\end{equation}
where $ \varUpsilon_{Q} $ and $ \varUpsilon_{\rm U} $ denote the 
contribution of intrinsic degree of freedom $ Q $ and of Unruh effect, 
respectively. The corresponding quantities are calculated in the 
Appendix; analytic expressions for the terms are presented by 
Eq.\eqref{varUpsilon Q} and Eq.\eqref{varupsilon U 2}, respectively.

\begin{figure}
        \centering
        \includegraphics[
                        trim={60 0 40 0},
                        clip,
                        width=0.45\textwidth
                ]{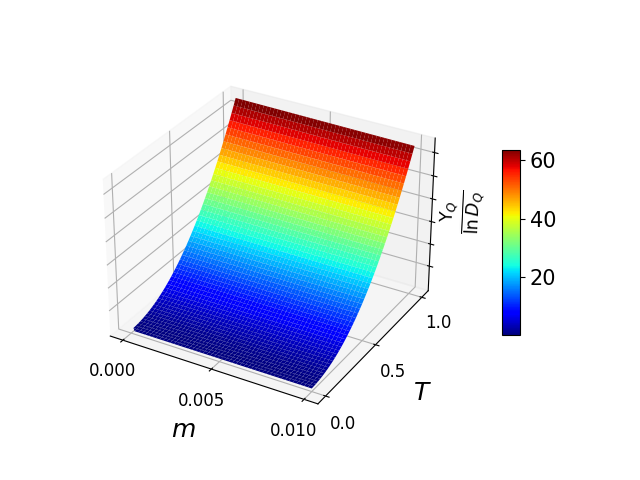}
        \includegraphics[
                        trim={60 0 40 0},
                        clip,
                        width=0.45\textwidth
                ]{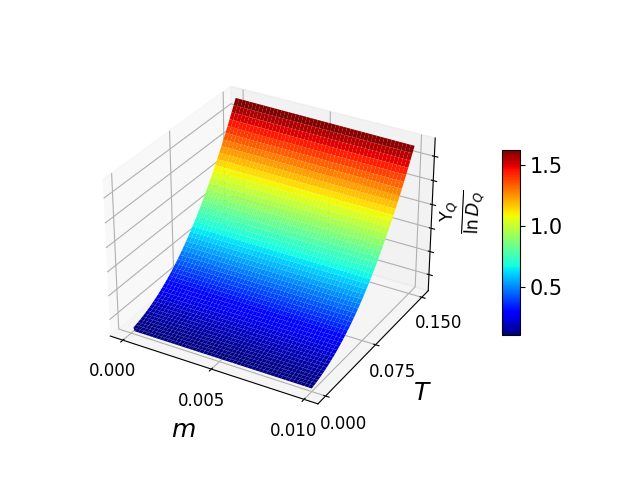}
        \caption{Left panel: quantity $ \varUpsilon_{Q}/\ln D_{Q} $ 
from Eq.\eqref{varUpsilon Q} as a function of $ m $ and $ T $. 
Right panel: the same as the left one but for the low temperature 
range.}
        \label{fig:q-upsilon}
\end{figure}

The scaled term $ \varUpsilon_{Q} $ depends on two parameters: the 
mass $ m $ of emitted particles and the horizon temperature $ T $. It 
is depicted in Fig.~\ref{fig:q-upsilon}. As one can see, the quantity 
gradually increases with temperature, in full accord with 
Eq.\eqref{H_U(n|N,E/T) asymptotics}. For hot horizons, when 
$ T \gtrsim 0.1 $, the ratio $ \varUpsilon_{Q}/\ln D_{Q} $ exceeds 
unity. Recall, that Eq.\eqref{ratio} is measured in the units of 
$ H_{\rm BH} $, so breaking the threshold determines the applicability 
of the imposed model restrictions.

\begin{figure}
        \centering
        \includegraphics[
                        trim={60 0 40 0},
                        clip,
                        width=0.45\textwidth
                ]{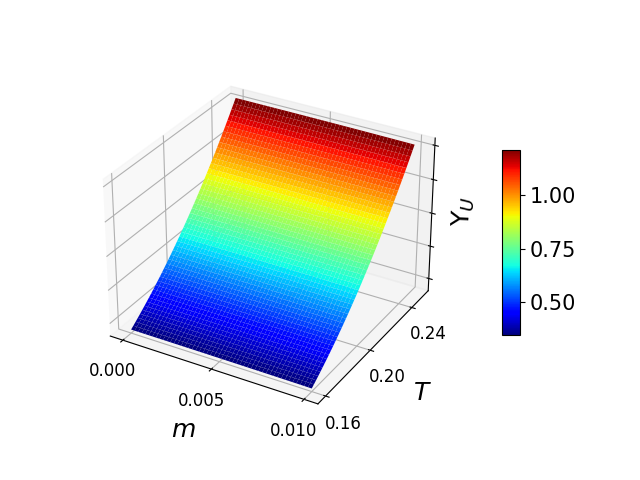}
        \includegraphics[
                        trim={60 0 40 0},
                        clip,
                        width=0.45\textwidth
                ]{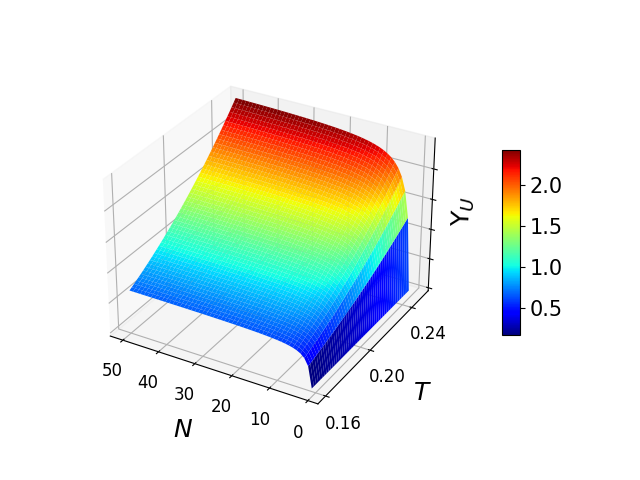}
        \caption{Left panel: the term $ \varUpsilon_{\rm U} $ from 
Eq.\eqref{varupsilon U 2} as a function of $ m $ and $T$ for $N = 2$. 
Right panel: the same term as a function of $N$ and $T$ at $ m = 0 $.}
        \label{fig:u-upsilon}
\end{figure}

Contrary to the first term at the \emph{rhs} of Eq.\eqref{ratio}, the 
second one turns out to be also a function of $ N $. 
Figure~\ref{fig:u-upsilon} displays the dependence of 
$ \varUpsilon_{\rm U} $ on $ m,T $ for $ N = 2 $ (left panel) and on 
$ N,T $ for massless particles (right panel). Again, the contribution 
of Unruh radiation to the black hole entropy $ H_{\rm BH} $ gradually 
increases with temperature and $ N $. The observed truncation of the 
plot at $ T \lesssim 0.16 $ is caused by the numerical precision. This 
result can be easily deduced by analytic estimates of 
Eq.\eqref{varupsilon U 2}; see also Eq.\eqref{H_U(n|N,E/T) asymptotics} 
and the text therein.

\section{Asymptotic analysis}
\label{sec:analysis}

As follows from Eq.\eqref{varUpsilon Q} and Eq.\eqref{varupsilon U 2}, 
for hot horizons the ratio given by Eq.\eqref{ratio} quickly rises up 
with increasing $ T $. For Planck temperature $ T = 1 $ it may easily 
exceed unity. Thus, the proposed model is not valid when $T \approx 1$. 
It can be argued that this is due to the insensitivity of the emission 
probability to energy. However, the exact energy distribution 
$ \left\{E\right\} $ can be determined from the state hidden below the 
event horizon, and cannot be treated without gravity quantization.

For the black hole of stellar scale the situation is different. In the 
case $ T \to 0 $ the Unruh term $ \varUpsilon_{\rm U} $ vanishes, and 
$ \varUpsilon_{Q} $ dominates, as seen from Eq. \eqref{ratio}, 
Eq.\eqref{varUpsilon Q} and Eq.\eqref{varupsilon U 2}:
\begin{equation}\label{H_BH T -> 0}
    \lim\limits_{T \to 0}\frac{H_{M}\left(Q,n,E|N,T\right)}{H_{\rm BH}}
		= \frac{1 - 3m^2 + 2m^3}{3\pi} \ln D_Q
		\geq \frac{\ln D_Q}{3\pi}\ .
\end{equation}

\begin{figure}
	\centering
	\includegraphics[width=0.7\textwidth]{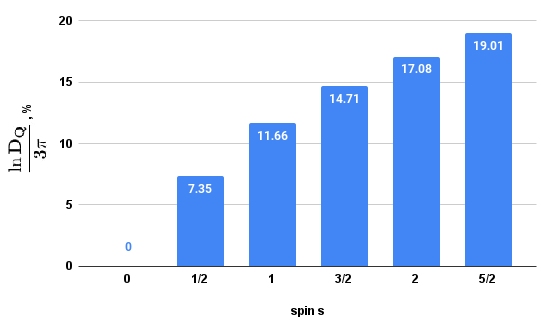}
	\caption{Lower bound for 
$ \lim\limits_{T \to 0}\frac{H_{M}\left(s,n,E|T\right)}{H_{\rm BH}} $ 
        at different values of spin.}
	\label{fig:lower bound}
\end{figure}

The lower bound of Eq.\eqref{H_BH T -> 0} for the spin degree of 
freedom is shown in Fig.~\ref{fig:lower bound}. Total Unruh entropy 
gradually increases with spin. Here the case $ s = 0 $ corresponds to 
the probability of particle emission at zeroth temperature, when the 
source cannot emit thermal radiation at all. This result is in line 
with the previous studies \cite{a_bh_entr_en,a_bh_entr_en_small}. 
However, even for $ s = 1/2 $ the ratio reaches $ 7.35\% $, and for 
$ s = 5/2 $ it touches $ 19.01\% $.

As follows from Eq.\eqref{varUpsilon Q} and Eq.\eqref{varupsilon U 2}, 
massive particles decrease the total Unruh entropy 
$ H_{M}\left(Q,n,E|N,T\right) $. Some terms survive even in the case 
of small temperatures, see Eq.\eqref{H_BH T -> 0}. This hints the 
influence of correlations in the outgoing radiation, which are encoded 
by massive particles only. One can consider the effect as some 
information outflow from the system \cite{bhi_Hawking_2,bhe_ev} within, 
e.g., the Page formalism \cite{PhysRevLett.71.1291,PhysRevLett.71.3743}. 
On the other hand, the decrease might originate from the black hole's 
energy, namely, the more of it is deposited in $ m $, the less can be 
used for the entropy production. The effect is insignificant due to 
the smallness of $ m $, since $ m \ll 1 $ for any elementary particles. 
Note also that the lower bound in Eq.\eqref{H_BH T -> 0} is 
non-negative for any values of $ m $.
It should be mentioned that the term surviving in the limit $ T \to 0 $ 
is the only one obeying the area law.

\section{Conclusions}
\label{sec:conclusion}
 
We calculated the total entropy of Unruh radiation from a Schwarzschild 
black hole of mass $ M $. To estimate the contribution of $ D = 1 + 1 $ 
effect in $ D = 3 + 1 $ space-time, the black hole was represented as a 
set of Unruh horizons. The treatment is valid in the case of 
quasiclassical approach, when any back-reactions and quantum gravity 
effects are insignificant. The method allows us to take into account 
all intrinsic degrees of freedom of emitted particles, such as spin 
$ s $, charge etc., together with their mass $ m $ and multiplicity,
defined by the parameter $ N $.

As far as information theory is concerned, the Unruh entropy is nothing 
but the result of inaccessibility to the whole state. This has much in 
common with the case of limited (partial) access to some chain of bits 
(qubits), in full accord with the Page formalism. Due to its geometric 
treatment, the approach implies a unitary evolution for gravity and 
obeys the information conservation.

The calculated entropy $ H_{M}\left(s,n,E|N,T\right) $ is represented 
in exact analytic form. It contains the terms obeying the area law. 
Surprisingly enough, these terms are governed not by the Unruh effect 
but by quantum numbers of outgoing particles only.

The entropy $ H_{M}\left(q,n,E|N,T\right) $ contains also negative 
terms, which are proportional to powers of $ m $ and do not vanish even 
for $ T \to 0 $. The role of this effect is still unclear. This can be 
interpreted as some information outflow from the BH or just a simple 
consequence of energy depositing by massive particles.

The intrinsic degrees of freedom have a significant impact on the 
entropy, in full accord with the growth of phase space for the Unruh 
radiation. The ratio Eq.\eqref{ratio} may exceed unity if the relevant 
Hilbert space is large enough, thus imposing restrictions to the model.

However, this does not mean that the use of the Unruh effect is 
incorrect for the entropy problem. Recall that the additive entropy 
behavior is violated by conservation laws. The emerging correlations 
will induce conditional distributions and, therefore, lower the joint 
entropy due to Eq.\eqref{H conditional}. For example, spin spectra are 
not independent because of momentum conservation. Particles with 
different spins will be correlated, thus resulting in the non-extensive 
behavior of joint entropy.

Note that energy-induced correlations can exactly reproduce the area 
law, thus favoring evaporation at the event horizon to be unitary
\cite{bhe_cond}. In support of the idea, recent study 
\cite{CALMET2023137820} shows that quantum hair is crucial for the 
emission spectrum of a black hole. Therefore, the proper 
analysis of relevant conditional distributions is of great importance 
for the black hole entropy problem and deserves further investigation.

\acknowledgments

M.T. acknowledges financial support of the Norwegian Directorate for 
Higher Education and Skills (DIKU) under Grant
``CPEA-LT-2016/10094 - From Strong Interacting Matter to Dark Matter."
The work of L.B. and E.Z. was supported by the Norwegian Research 
Council (NFR) under grant No. 255253/F50 - “CERN Heavy Ion Theory”.
Computer calculations were made at SAGA (UiO, Oslo)
computer cluster facilities.

\appendix
\section{Analytic expressions for $ \varUpsilon_{Q} $ and 
$ \varUpsilon_{\rm U} $}

The term $ \varUpsilon_{Q} $ can be calculated analytically:
\begin{equation}\label{varUpsilon Q}
\begin{split}
    \varUpsilon_{Q}
	&= 4\pi T^2
	   \int_{m}^{1}
             \left(\sqrt{\frac{E^2-m^2}{4\pi^2T^2} + 1} + 1\right)^2
	       	{\rm d}E \ln D_Q\\
		&= \Biggl[
			\frac{1 - 3m^2 + 2m^3}{3\pi}
			  + 2T\sqrt{1+4\pi^2T^2-m^2}
			  + 4\pi T^2\left(1-2m\right)\\
		&\hspace{3em}
		 + 2T\left(4\pi^2T^2-m^2\right)
			\ln\frac{1+\sqrt{1+4\pi^2T^2-m^2}}{2\pi T +m}
			\Biggr]\ln D_Q\ .
\end{split}
\end{equation}

The term $ \varUpsilon_{\rm U} $, see Eq.\eqref{H_U(n|N,E/T)} and 
Eq.\eqref{D_a}, reads
\begin{equation}\label{varUpsilon U 1}
       \varUpsilon_{\rm U}
	  = 4\pi T^2
	     \int_{m}^{1}
		  \left(
			\sqrt{\frac{E^2-m^2}{4\pi^2T^2} + 1}
			  + 1
		            \right)^2
			\left[
				\sigma\left(E/T\right)
				  - \sigma\left(NE/T\right)
				\right]
				{\rm d}E\ .
\end{equation}
It contains incomplete Bose-Einstein integrals and can be calculated 
as follows.

Rewriting Eq.\eqref{sigma} as
\begin{equation*}
   \sigma(qE/T) = \sum_{k=1}^{\infty}\left(qE/T + 1/k\right)e^{-kqE/T}
\end{equation*}
and using lower incomplete gamma functions $ \gamma\left(\nu,x\right) $
\begin{equation*}
\gamma\left(\nu,x\right) = \int_{0}^{x}t^{\nu-1}e^{-t}{\rm d}t = 
\left(\nu-1\right)!\left(1 - e^{-x}\sum_{j=0}^{\nu-1}\frac{x^j}{j!}
                                 \right)\ ,
\end{equation*}
we obtain
\begin{equation}\label{integral E^nu sigma}
\begin{split}
\int_{m}^{1}\sigma\left(qE/T\right)E^\nu{\rm d}E
	= \frac{T^{\nu+1}}{q^{\nu+1}}
		\sum_{k=1}^{\infty}
			\frac{\gamma\left(\nu+1,x\right)
		  + \gamma\left(\nu+2,x\right)}{k^{\nu+2}}
			\Bigg\vert_{x=kqm/T}^{x=kq/T}\ .
\end{split}
\end{equation}
Here and below we use the notation
\begin{equation}\label{f(a) - f(b)}
	f\left(x\right)\Big|_{x=b}^{x=a}
		= f\left(a\right)
		- f\left(b\right)\ .
\end{equation}

Then, substituting Eq.\eqref{integral E^nu sigma} into 
Eq.\eqref{varUpsilon U 1} and using the decomposition
\begin{equation*}
   \left(1+x\right)^{\alpha}
     = \sum_{n=0}^{\infty}\binom{\alpha}{n}x^n\ ,
       \quad \left|x\right|<1\ ,
\end{equation*}
we obtain that
\begin{equation}\label{varupsilon U 2}
\begin{split}
	\varUpsilon_{\rm U}
		&= \frac{T}{\pi}
		   \Biggl[
				\left(8\pi^2T^2-m^2\right)\\
		&\hspace{3em}
		  \times\sum_{k=1}^{\infty}
	  \frac{\gamma\left(1,x\right) + \gamma\left(2,x\right)}{k^2}
				\left(
			  \Bigl.\Bigr|_{x=km/T}^{x=k/T}
			- \frac{1}{N}\Bigl.\Bigr|_{x=kNm/T}^{x=kN/T}
				\right)\\
		&\hspace{3em}
		  + T^2\sum_{k=1}^{\infty}
			\frac{\gamma\left(3,x\right)
				+ \gamma\left(4,x\right)}{k^{4}}
					\Biggl(
			  \Bigl.\Bigr|_{x=km/T}^{x=k/T}
			- \frac{1}{N^3}\Bigl.\Bigr|_{x=kNm/T}^{x=kN/T}
						\Biggr)\\
		&\hspace{3em}
		  + 8\pi^2T^2
			\sum_{n=0}^{\infty}
			 	\binom{1/2}{n}
			 	\begin{cases}
	             A_{1},	&	2\pi T > \sqrt{1-m^2}\\
		A_{\mu} + B,	&	2\pi T < \sqrt{1-m^2}
				\end{cases}
		   \Biggr]\ ,
\end{split}
\end{equation}
where
\begin{equation*}
\begin{split}
	A_{\beta}
		&= \frac{1}{\left(2\pi\right)^{2n}}
		   \sum_{q=0}^{n}
			\binom{n}{q}
			\left(-1\right)^q
			\frac{m^{2q}}{T^{2q}}
			\sum_{k=1}^{\infty}
   		   	\frac{\gamma\left(1+2n-2q,x\right)
   			+\gamma\left(2+2n-2q,x\right)}{k^{2+2n-2q}}\\
		&\quad\times
			\left(
				\Bigl.\Bigr|_{x=km/T}^{x=k\beta/T}
			  - \frac{1}{N^{1+2n-2q}}
			    \Bigl.\Bigr|_{x=kNm/T}^{x=kN\beta/T}
			\right)\\
	B	&= \left(2\pi\right)^{2n-1}
		   \sum_{q=0}^{\infty}
		   	\binom{\frac{1}{2}-n}{q}
		   	\left(-1\right)^q
		   	\frac{m^{2q}}{T^{2q}}
		   	\sum_{k=1}^{\infty}
	   		\frac{\gamma\left(2-2n-2q,x\right)
   			+ \gamma\left(3-2n-2q,x\right)}{k^{3-2n-2q}}\\
		&\quad
		  \times\left(
		  			\Bigl.\Bigr|_{x=k\mu/T}^{x=k/T}
		  		  - \frac{1}{N^{2-2n-2q}}
		  		    \Bigl.\Bigr|_{x=kN\mu/T}^{x=kN/T}
		  		\right)\\
	\mu &= \sqrt{4\pi^2 T^2 + m^2}\ .
\end{split}
\end{equation*}


\begin{thebibliography}{99}

\bibitem{bhi_Bekenstein}
J.~D.~Bekenstein,
\emph{Black holes and entropy},
\emph{Phys. Rev. D} {\bf 7} (1973) 2333

\bibitem{bhi_4_laws}
J.~M.~Bardeen, B.~Carter and S.~W.~Hawking,
\emph{The four laws of black hole mechanics},
\emph{Commun. Math. Phys.} {\bf 31} (1973) 161

\bibitem{bhi_Hawking}
S.~W.~Hawking,
\emph{Particle creation by black holes},
\emph{Commun. Math. Phys.} {\bf 43} (1975) 199

\bibitem{PhysRevD.14.2460}
S.~W.~Hawking,
\emph{Breakdown of predictability in gravitational collapse},
\emph{Phys. Rev. D} {\bf 14} (1976) 2460

\bibitem{PhysRevLett.80.904}
A.~Ashtekar, J.~Baez, A.~Corichi and K.~Krasnov,
\emph{Quantum geometry and black hole entropy},
\emph{Phys. Rev. Lett.} {\bf 80} (1980) 904

\bibitem{bhl_Khriplovich_1}
I.~B.~Khriplovich,
\emph{Entropy and area of black holes in loop quantum gravity},
\emph{Phys. Lett. B} {\bf 537} (2002) 125

\bibitem{bhl_e_spectr}
I.~B.~Khriplovich,
\emph{Holographic bound and spectrum of quantized black hole},
arXiv:gr-qc/0411109

\bibitem{bhe_Immirzi}
T.~Jacobson,
\emph{A note on renormalization and black hole entropy in loop quantum 
gravity},
\emph{Class. Quantum Grav.} {\bf 24} (2007) 4875

\bibitem{bhl_su2}
E.~Bianchi,
\emph{Black hole entropy, loop gravity, and polymer physics},
\emph{Class. Quantum Grav.} {\bf 28} (2011) 114006

\bibitem{bhl_intertwiner}
E.~R.~Livine and D.~R.~Terno,
\emph{Entropy in the classical and quantum polymer black hole models},
\emph{Class. Quantum Grav.} {\bf 29} (2012) 224012

\bibitem{bhef_Verlinde}
E.~Verlinde,
\emph{On the origin of gravity and the laws of Newton},
\emph{J. High Energ. Phys.} {\bf 04} (2011) 029

\bibitem{bhe_str}
G.~T.~Horowitz,
\emph{Black holes, entropy, and information}, 
arXiv:0708.3680 [astro-ph]

\bibitem{bhe_micro}
A.~Strominger and C.~Vafa,
\emph{Microscopic origin of the Bekenstein-Hawking entropy},
\emph{Phys. Lett. B} {\bf 379} (1996) 99

\bibitem{bhe_M-th}
R.~Emparan and G.~T.~Horowitz,
\emph{Microstates of a neutral black hole in M theory},
\emph{Phys. Rev. Lett.} {\bf 97} (2006) 141601

\bibitem{PhysRevD.75.084006}
R.~Emparan and A.~Maccarrone,
\emph{Statistical description of rotating Kaluza-Klein black holes},
\emph{Phys. Rev. D} {\bf 75} (2007) 084006

\bibitem{bhe_M-th_Kerr}
G.~T.~Horowitz and M.~M.~Roberts,
\emph{Counting the microstates of a Kerr black hole},
\emph{Phys. Rev. Lett.} {\bf 99} (2007) 221601

\bibitem{bhe_ev}
C.~A.~S.~Silva and R.~R.~Landim,
\emph{A note on black-hole entropy, area spectrum, and evaporation},
\emph{Europhys. Lett.} {\bf 96} (2011) 10007

\bibitem{bhi_Hawking_2}
S.~W.~Hawking,
\emph{Information loss in black holes},
\emph{Phys. Rev. D} {\bf 72} (2005) 084013

\bibitem{1403.7314}
B.~Zhang, Q.~Cai, M.~Zhan and L.~You,
\emph{Correlation, entropy, and information transfer in black hole
      radiation},
\emph{Chi. Sci. Bull.} {\bf 59} (2014) 1057

\bibitem{bhe_Hooft}
G.~'t~Hooft,
\emph{On the quantum structure of a black hole},
\emph{ Nucl. Phys. B} {\bf 256} (1985) 727

\bibitem{bhe_bw_unc}
R.~Brustein and J.~Kupferman,
\emph{Black hole entropy divergence and the uncertainty principle},
\emph{Phys. Rev. D} {\bf 83} (2011) 124014

\bibitem{bhe_bw_cor}
K.~Wontae and K.~Shailesh,
\emph{Higher order WKB corrections to black hole entropy in brick wall
      formalism},
\emph{Eur. Phys. J. C} {\bf 73} (2013) 2398

\bibitem{Almheiri_2013}
A.~Almheiri, D.~Marolf, J.~Polchinski and J.~Sully,
\emph{Black holes: complementarity or firewalls?}
\emph{J. High Energ. Phys.} {\bf 02} (2013)

\bibitem{https://doi.org/10.1007/s10701-017-0122-3}
G.~'t~Hooft,
\emph{The firewall transformation for black holes and some of its
      implications},
\emph{Found. Phys.} {\bf 47} (2017) 1503

\bibitem{bhef_holo}
L.~Susskind,
\emph{The world as a hologram},
\emph{J. Mathem. Phys.} {\bf 36} (1995) 6377

\bibitem{bhef_en}
S.~Ryu and T.~Takayanagi,
\emph{Holographic derivation of entanglement entropy from the 
      anti--de Sitter space / conformal field theory correspondence},
\emph{Phys. Rev. Lett.} {\bf 96} (2006) 181602

\bibitem{bhef_en_ext}
S.~Ryu and T.~Takayanagi,
\emph{Aspects of holographic entanglement entropy},
\emph{J. High Energ. Phys.} {\bf 08} (2006) 045

\bibitem{1108.2650}
A~Davidson,
\emph{Holographic shell model: Stack data structure inside black holes},
\emph{ Int. J. Mod. Phys. D} {\bf 23} (2014) 1450041

\bibitem{bhe_Srednicki}
M.~Srednicki,
\emph{Entropy and area},
\emph{Phys. Rev. Lett.} {\bf 71} (1993) 666

\bibitem{bhe_dof_rev}
S.~Das, S.~Shankaranarayanan, and S.~Sur,
\emph{Black hole entropy from entanglement: A review}, 
arXiv:0806.0402 [gr-qc]

\bibitem{bhe_en_short}
A.~Iorio, G.~Lambiase and G.~Vitiello,
\emph{Quantization of scalar fields in curved background, deformed Hopf
      algebra and entanglement},
arXiv:quant-ph/0207173

\bibitem{bhe_en_2}
A.~Iorio, G.~Lambiase and G.~Vitiello,
\emph{Entangled quantum fields near the event horizon and entropy},
\emph{Ann. Phys. (NY)} {\bf 309} (2004) 151

\bibitem{bhe_dof}
S.~Das and S.~Shankaranarayanan,
\emph{Where are the black-hole entropy degrees of freedom?},
\emph{Class. Quantum Grav.} {\bf 24} (2007) 5299

\bibitem{bhe_atm}
T.~Jacobson and R.~Parentani,
\emph{Black hole entanglement entropy regularized in a freely falling
      frame},
\emph{Phys. Rev. D} {\bf 76} (2007) 024006

\bibitem{mathur2023universality}
S.D.~Marthur and M.~Mehta,
\emph{The universality of black hole thermodynamics},
arXiv:2305.12003 [hep-th]

\bibitem{bht_rev}
S.~Carlip,
\emph{Black Hole Thermodynamics and Statistical Mechanics},
Springer, Berlin (2009) pp. 89--123

\bibitem{bht_rev_2}
D.~N.~Page,
\emph{Hawking radiation and black hole thermodynamics},
\emph{New J. Phys.} {\bf 7} (2005) 203

\bibitem{1807.05864}
O.~C.~Stoica,
\emph{Revisiting the black hole entropy and the information paradox},
\emph{Adv. High En. Phys.} {\bf 2018} (2018) 4130417

\bibitem{1804.10610}
A.~C.~Wall,
\emph{A survey of black hole thermodynamics}, 
arXiv:1804.10610 [gr-qc]

\bibitem{Polchinski_2016}
J.~Polchinski,
\emph{The black hole information problem},
\emph{New Frontiers in Fields and Strings} 
World Scientific, Singapore (2017) pp. 353--397

\bibitem{Unruh_2017}
W.~G.~Unruh and R.~M~Wald,
\emph{Information loss},
\emph{Rep. Progr. Phys.} {\bf 80} (2017) 092002

\bibitem{RevModPhys.88.015002}
D.~Harlow,
\emph{Jerusalem lectures on black holes and quantum information},
\emph{Rev. Mod. Phys.} {\bf 88} (2016) 015002

\bibitem{u_Unruh}
W.~G.~Unruh,
\emph{Notes on black-hole evaporation},
\emph{Phys. Rev. D} {\bf 14} (1976) 870

\bibitem{RevModPhys.82.277}
J.~Eisert, M.~Cramer and M.~B.~Plenio,
\emph{Colloquium: Area laws for the entanglement entropy},
\emph{Rev. Mod. Phys.} {\bf 82} (2010) 277

\bibitem{RevModPhys.80.787}
L.~C.~B.~Crispino, A.~Higuchi and G.~E.~A.~Matsas,
\emph{The Unruh effect and its applications},
\emph{Rev. Mod. Phys.} {\bf 80} (2008) 787

\bibitem{0903.0250}
R.~Banerjee and B.~R.~Majhi,
\emph{Hawking black body spectrum from tunneling mechanism},
\emph{Phys. Lett. B} {\bf 675} (2009) 243

\bibitem{0908.3149}
D.~Roy,
\emph{The Unruh thermal spectrum through scalar and fermion tunneling},
\emph{Phys. Lett. B} {\bf 681} (2009) 185

\bibitem{particles5020014}
M.~Teslyk, O.~Teslyk, L.~Zadorozhna, L.~Bravina and E.~Zabrodin,
\emph{Unruh effect and information entropy approach},
\emph{Particles} {\bf 5} (2022) 157

\bibitem{pathak} A.~Pathak,
\emph{Elements of Quantum Computation and Quantum Communication},
Taylor \& Francis, London (2013) pp. 1--340

\bibitem{PhysRevD.7.2850}
S.~A.~Fulling,
\emph{Nonuniqueness of canonical field quantization in Riemannian
      space-time},
\emph{Phys. Rev. D} {\bf 7} (1973) 2850

\bibitem{10.1088/0305-4470/8/4/022}
P.~C.~W.~Davies,
\emph{Scalar production in Schwarzschild and Rindler metrics},
\emph{J. Phys. A} {\bf 8} (1975) 609

\bibitem{a_bh_entr_en}
E.~D.~Belokolos and M.~V.~Teslyk,
\emph{Scalar field entanglement entropy of a Schwarzschild black hole
      from the Schmidt decomposition viewpoint},
\emph{Class. Quantum Grav.} {\bf 26} (2009) 235008

\bibitem{a_bh_entr_en_small}
M.~V.~Teslyk and O.~M.~Teslyk,
\emph{Scalar field entanglement entropy for a small Schwarzschild black
      hole},
\emph{Class. Quantum Grav.} {\bf 30} (2013) 125013

\bibitem{PhysRevLett.71.1291}
D.~N.~Page,
\emph{Average entropy of a subsystem},
\emph{Phys. Rev. Lett.} {\bf 71} (1993) 1291

\bibitem{PhysRevLett.71.3743}
D.~N.~Page,
\emph{Information in black hole radiation},
\emph{Phys. Rev. Lett.} {\bf 71} (1993) 3743

\bibitem{bhe_cond}
Baocheng~Zhang, Qing~yu~Cai, Li~You and Ming~sheng~Zhan,
\emph{Hidden messenger revealed in Hawking radiation: A resolution to 
      the paradox of black hole information loss},
\emph{Phys. Lett. B} {\bf 675} (2009) 98

\bibitem{CALMET2023137820}
X.~Calmet, S.~D.~H.~Hsu and M.~Sebastianutti,
\emph{Quantum gravitational corrections to particle creation by black
      holes},
\emph{Phys. Lett. B} {\bf 841} (2023) 137820

\end{thebibliography}
\end{document}